\documentclass{article}
\usepackage[latin9]{inputenc}
\usepackage{color}
\usepackage{array}
\usepackage{verbatim}
\usepackage{multirow}
\usepackage{amsmath}
\usepackage{amssymb}
\usepackage{graphicx}
\usepackage{subfigure}

\makeatletter

\usepackage{spconf}

\begin{document}

\title{Vectors of Locally Aggregated Centers \\ for Compact Video Representation}

%
\name{Alhabib Abbas, Nikos Deligiannis and Yiannis Andreopoulos\thanks{This work appeared in the Proc. of the IEEE Int. Conf. on Multimedia and Expo, ICME\ 2015, Torino, Italy. The work was funded in part by Innovate UK, project REVQUAL - Resolving Visual Quality for Media (101855).}}
\address{Department of Electronic and Electrical Engineering\\
University College London (UCL), London, U.K.\\
\{alhabib.abbas.13, n.deligiannis, i.andreopoulos\}@ucl.ac.uk}

\maketitle

\begin{abstract}
We propose a novel vector aggregation technique for compact video
representation, with application in accurate similarity detection
within large video datasets. The current state-of-the-art in visual  search  is formed by the vector of locally aggregated descriptors (VLAD) of Jegou
\emph{et al.} VLAD  generates  compact video representations
based on scale-invariant feature transform (SIFT)  vectors (extracted per frame) and local feature centers computed over
a training set. With the aim to increase robustness to visual distortions, we propose a new approach that operates at a coarser level in the feature
representation.  We create \emph{vectors of locally aggregated
centers} (VLAC)  by first clustering SIFT features to obtain   \emph{local feature centers} (LFCs)  and then encoding the latter with respect to given centers of local feature centers (CLFCs), extracted from a training
set. The sum-of-differences between the LFCs and the CLFCs are aggregated  to generate an extremely-compact video description
 used for accurate video segment similarity detection.
Experimentation using  a video dataset, comprising
more than 1000 minutes of  content from the
Open Video Project, shows that  VLAC  obtains substantial gains in terms of mean
Average Precision (mAP) against VLAD and the hyper-pooling  method
of Douze \emph{et al.},  under the same compaction factor and the
same set of  distortions.   
\end{abstract}


\section{Introduction}

\label{sec:intro}

Recommendation services, event detection, clustering and categorization
of video data, and retrieval algorithms for large video databases depend on efficient and reliable similarity identification amongst video segments  \cite{arandjelovic2013all, chou2013near,wang2010large}. In a nutshell, given a query video, we wish to find all similar video segments within a large video database in the most reliable and efficient way. The state-of-the-art in similarity identification hinges on video fingerprinting algorithms \cite{revaud2013event, douze2013stable}. The aim of such algorithms is to provide for distinguishable representations that remain robust under visual distortions, such as, rotation, compression, blur, resizing, flicker, etc. Such distortions are expected to be present within large video collections, or when dealing with content {\textquotedblleft \textit{in the wild}\textquotedblright} \cite{wang2010large}. 

In a broad sense, video similarity identification can be seen as a spatio-temporal matching problem via an appropriate feature space or descriptor. Recent results  have shown that similarity identification algorithms based on local descriptors, such as the scale invariant feature transform (SIFT) \cite{lowe1999object} or dense SIFT \cite{vedaldi2010vlfeat}, tend to significantly outperform previous approaches based on histogram methods \cite{hampapur2001comparison} or fingerprinting algorithms \cite{esmaeili2011robust}, especially under the presence of distortions in the video data. Therefore, the state-of-the-art in this area is based on vectors of locally aggregated descriptors (VLAD) \cite{jegou2012aggregating}, or Bag-of-Words (BoW) methods \cite{yang2007evaluating}, which merge feature descriptors in video frame regions. More recently, hyper-pooling approaches have been proposed \cite{douze2013stable}, which perform two consecutive VLAD stages in order to compact entire video sequences into a unique aggregated descriptor vector. 

In this paper, we focus on VLAD-based algorithms and examine the problem
of creating compact representations that are suitable for efficient
and accurate similarity identification of segments of videos within
a large video collection. The paper makes the following contributions: 
\begin{itemize}
\item Instead of creating holistic hyper-pooling approaches for entire
video sequences, we concentrate on groups of frames (GoFs) within
a video sequence in order to allow for video segment search. 
\item Instead of directly compacting  feature descriptors, we follow a two-stage clustering approach: we first cluster features to obtain local-feature-centers (LFCs)  and then encode the latter with respect to a given set of 
centers of local-feature-centers (CLFCs),  computed from a training set. 
\item Similar to  VLAD,
 we encode the LFCs by aggregating their differences with respect to their
corresponding CLFCs, thereby creating \emph{vectors of locally aggregated
centers} (VLAC). 
\item Experiments using a 100-minute  training  set and a 1000-minute
test set from the Open Video Project reveal that, for the same compaction
factor, our proposal is outperforming the state-of-the-art VLAD method \cite{jegou2012aggregating} by more than  15\% in terms of mean Average Precision (mAP). 
\end{itemize}
The remainder of the paper is  as follows. Section \ref{sec:Background}
summarizes the operation of VLAD and hyper-pooling  that constitute the state-of-the-art and form the basis of the proposed compaction algorithm.
Section \ref{sec:VLAC} presents the proposed VLAC approach. Section
\ref{sec:Evaluation} presents experimental results, while  Section \ref{sec:Conclusion}
draws concluding remarks.

\section{Background on VLAD and Hyper-Pooling}
\label{sec:Background}

\subsection{Visual Feature Description}

Current solutions make use of image descriptors to represent individual
frames within a video \cite{revaud2013event, douze2013stable}. After extracting the local feature descriptors of a given set of frames using an algorithm such as SIFT \cite{lowe1999object} or dense SIFT �\cite{vedaldi2010vlfeat}, these descriptors are then accumulated
to produce a compact frame representation. Recent work advocated the
use of pooling strategies instead of simple averaging methods, in
order to minimize information loss. A common way to achieve this is
by using BoW methods \cite{yang2007evaluating} or VLAD \cite{jegou2012aggregating}. In this paper, we focus on the latter
as it has been shown to achieve state-of-the-art results in terms
of mAP in medium and large-scale sets of image and video content.

\subsection{Vector of Locally Aggregated Descriptors}

VLAD \cite{arandjelovic2013all, jegou2012aggregating} is a vector aggregation algorithm that produces a fixed-size
compact description of a set comprising a variable number of data
points. VLAD was proposed as a novel approach aimed to optimize:
\textit{(i)} the representation of aggregated feature descriptors;
\textit{(ii)} the dimensionality reduction;
\textit{(iii)} the indexing of the output vectors.

These aspects are interrelated---for example, dimensionality reduction
directly affects the way we index the output vectors. While high dimensional
vectors produce more accurate search results, low dimensional
vectors are easier to index and require less operations and storage.

Consider a set of $W$ video frames to be used for training purposes.
For the $w$th training frame, $1\leq w\leq W$, a visual feature
detector and descriptor (e.g., the SIFT detector and descriptor \cite{lowe1999object})
is calculated, thereby producing $K_{w}$ 
feature vectors $\mathbf{f}_{w,k}$, $1\leq k\leq K_{w}$, each with dimension $1\times F$. The ensemble
of these features comprises the $w$th training frame's set of visual
features $\mathfrak{\mathcal{F}}_{w}=\left\{ \mathbf{f}_{w,1},\,\mathbf{f}_{w,2},...,\,\mathbf{f}_{w,K_{w}}\right\} $.
The concatenation of all these sets for all $W$ training frames,
given by $\mathcal{F}_{\text{train}}=\left\{ \mathcal{F}_{1},\,\mathcal{F}_{2},...,\,\mathcal{F}{}_{W}\right\} $,
undergoes a clustering approach, such as K-means \cite{bishop2006pattern}, thereby
grouping all vectors in $\mathfrak{\mathcal{F}}_{\text{train}}$ into
$J$ clusters, with centers denoted by set $\mathcal{C}_{\text{train}}=\{\mathbf{c}_{1},\,\mathbf{c}_{2},...,\,\mathbf{c}_{J}\}$.
VLAD then encodes the set of visual features, $\mathfrak{\mathcal{F}}_{w}$,
of the $w$th frame as the group of $F$-dimensional vectors $\mathbf{v}_{w,j}$
($1\leq j\leq J$) given by
\begin{equation}
\mathbf{v}_{w,j}=\sum_{\forall k:\: Q\left(\mathbf{f}_{w,k}\right)=\mathbf{c}_{j}}\left(\mathbf{f}_{w,k}-\mathbf{c}_{j}\right)\label{eq:VLAD_center_coding}
\end{equation}
where $Q\left(\mathbf{f}_{w,k}\right)$ is the quantization function
that determines which cluster $\mathbf{f}_{w,k}$ belongs to.
Then, the VLAD of the $w$th frame is given by the 
vector of aggregated local differences
$\mathbf{v}_{w}=\begin{bmatrix}\mathbf{v}_{w,1} & \cdots & \mathbf{v}_{w,J}\end{bmatrix}$,  with dimension $1\times JF$.
All these vectors are concatenated into the $W\times JF$-dimensional
matrix $\mathbf{V}_{\text{train}}=\begin{bmatrix}\mathbf{v}_{1} & \cdots & \mathbf{v}_{W}\end{bmatrix}^{\text{T}}$,
which comprises the VLAD encoding of the training set. In order to
allow for further dimensionality reduction (thereby accelerating the
matching process), principal component analysis (PCA) is applied to $\mathbf{V_{\text{train}}}$,
and the $D$ most dominant eigenvectors are maintained in the $D\times JF$
matrix $\mathbf{P}_{\text{train}}$ in order to be used in the test
set.

When considering a test video frame, once its set of visual features
$\mathfrak{\mathcal{F}}_{\text{test}}=\left\{ \mathbf{f}_{\text{test},1},\,\mathbf{f}_{\text{test},2},...,\,\mathbf{f}_{\text{test},K_{\text{test}}}\right\}$
is produced by the SIFT descriptor (assuming $K_{\text{test}}$ points
were detected), VLAD performs the following step: \emph{(i)} calculation
of $\mathbf{v}_{\text{test},j}$ ($1\leq j\leq J$) via \eqref{eq:VLAD_center_coding}
with the precalculated center set $\mathcal{C}$; \emph{(ii)} aggregation
of these into a $JF\times1$ composite vector and application of dimensionality
reduction via the retained PCA coefficients in $\mathbf{P}_{\text{train}}$:
\begin{equation}
\mathbf{v}_{\text{test}}=\mathbf{P}_{\text{train}}\begin{bmatrix}\mathbf{v}_{\text{test},1} & \cdots & \mathbf{v}_{\text{test},J}\end{bmatrix},\label{eq:VLAD_coding}
\end{equation}
where $\mathbf{v}_{\text{test}}$ denotes the $D\times1$ VLAD of
the test video frame after compaction with PCA. The similarity between
two VLAD vectors of two test video frames $t_1$ and $t_2$
is simply measured via $s_{t_1,t_2}=\left\langle \mathbf{v}_{t_1},\mathbf{v}_{t_2}\right\rangle $.
Thresholding the set of similarity (i.e., inner product)\ results between a test video frame
and the entire test set of video frames provides the list of similar
frames retrieved under the selected threshold value.

\subsection{Hyper-Pooling}

A recent method proposed by Douze \emph{et al.}~\cite{douze2013stable} makes use of hyper-pooling (HP) strategies on the video description level.
Hyper-pooling works by using a second layer of data clustering and
encoding a set of frame VLAD descriptors into a single vector. 
Hyper-pooling utilizes an enhanced hashing scheme by exploiting the
temporal variance properties of  VLAD vectors \cite{douze2013stable} that have been produced per frame. After
performing  PCA, the temporal variance of VLAD vectors is
most prominent in the components associated with  low eigenvalues. Hence,  hyper-pooling postulates that we can get a more stable set of centers
by applying a clustering algorithm (such as K-means) on the set of
components relating to the highest eigenvalues. Indeed, hashing the
components that vary less with time has been shown to provide better
results in terms of stability and robustness to noise \cite{douze2013stable}.

\subsection{Motivation Behind the Proposed Concept\label{sec:Motivation}}

From the previous description, it is evident that the crucial aspects
of VLAD and hyper-pooling are the clustering and the PCA process performed
on the training set. Ideally, for a given set of video frames, we
would like to produce principal component vectors  for compaction of VLADs that do
not change substantially when the video frames undergo real-world visual distortions. For
example, consider two ensembles of training video frame sets, $\mathcal{I}_{\text{clean}}$
and $\mathcal{I}_{\text{noisy}}$, with the latter  produced by
distorting the video frames in  $\mathcal{I}_{\text{clean}}$ 
via blurring, compression artifacts, rotation, gamma changes, etc.
During the training stage, applying PCA on the  vectors of local differences (obtained per frame) will produce $D$ dominant eigenvectors forming the  $D\times JF$ matrices $\mathbf{P}_{\text{train,clean}}$
and $\mathbf{P}_{\text{train,noisy}}$. In case of  hyper-pooling the aforementioned matrices will have a dimension of $D\times JD_{0}$, where  $D_{0}$ is  the number of dimensions retained  after the first VLAD stage. Ideally,
the vectors in $\mathbf{P}_{\text{train,clean}}$ and $\mathbf{P}_{\text{train,noisy}}$
should be reasonably-well aligned, which is an indication that the
 compaction process is robust to noise. This can be tested by
computing the sum-of-inner-products between the $D$ dominant eigenvectors
of both cases: For both VLAD and hyper-pooling, we obtain 
\begin{align}
s_{\text{\{VLAD,HP\},clean,noisy}}&= \nonumber\\
&\sum_{i=1}^{D}\sum_{j=1}^{\left\{ JF,JD_{0}\right\} }p_{\text{train,clean}}\left[i,j\right]p_{\text{train,noisy}}\left[i,j\right]
\end{align}
where $p\left[i,j\right]$ denotes the $\left(i,j\right)$ element
of $\mathbf{P}$. 
We carried out such an indicative test in a set of $Z=2000$ video
frames taken from 10 video clips of 10-minute duration each. Each video underwent seven different
visual distortions, as tabulated in  Table \ref{tab:Distortions} and detailed in  Section \ref{sec:Evaluation}.
Using $J=128$ clusters for VLAD and
$F=128$ for dense SIFT,  we obtain $s_{\text{VLAD,clean,noisy}}=0.0085$
and $s_{\text{HP,clean,noisy}}=-0.0445$.  However,  utilizing the SIFT vectors directly, performing PCA decomposition to produce the two $D\times F$
matrices $\mathbf{P}_{\text{SIFT,train,clean}}$ and $\mathbf{P}_{\text{SIFT,train,noisy}}$,
and computing 
\begin{equation}
s_{\text{SIFT,clean,noisy}}=\sum_{i=1}^{D}\sum_{j=1}^{F}p_{\text{SIFT,train,clean}}\left[i,j\right]p_{\text{SIFT,train,noisy}}\left[i,j\right]
\end{equation}
we get $s_{\text{SIFT,clean,noisy}}=0.996$. The significant difference
between $s_{\text{SIFT,clean,noisy}}$ and $s_{\text{VLAD,clean,noisy}}$
and $s_{\text{HP,clean,noisy}}$ represents the reduction in tolerance
to distortions incurred when the vectors are projected to their principal components, which is performed in order  to gain the benefit of compaction. 

In this paper, our aim is to design a method leading to
the same compaction factor as VLAD, albeit having increased tolerance
to distortion in the video frames, which will allow for high recall
rates even when dealing with distorted versions of the input video
content. A secondary aim is to design our approach in a way that directly
deals  with video segments rather that individual video frames, thus allowing for video segment similarity detection. These two
aspects are elaborated in the next section.

\section{Vector of Locally Aggregated Centers\label{sec:VLAC}}
\label{sec:VLAC}

\subsection{VLAD per Video Frame}

The similarity between two videos can be estimated by obtaining the VLAD inner products per frame and averaging. We consider this approach as the baseline for video similarity
detection. This direct application of VLAD to video achieves good results
in terms of retrieval accuracy, albeit at the expense of high complexity
and storage requirements, even when the video is sampled at a substantially
lower frame-rate. All the solutions proposed are designed to approach
the performance of this baseline\  as much as possible while
requiring a fraction of its computational complexity and storage, or, alternatively, significantly-exceed the VLAD\ performance while incurring the same complexity and storage.

\subsection{Temporal Compaction for Video Segment Searching}

Video description algorithms such as hyper-pooling \cite{douze2013stable}  were designed for holistic
video description, namely,  the derived  vector  describes the entire video information \emph{as
a whole}. Temporal coherency is lost when using such holistic description
methods, thereby making the detection of video segments within longer
videos impossible. This problem can be solved by modifying holistic
solutions to work on \emph{groups of frames} (GoFs) within each given
video.  GoFs can be
viewed as fixed-size temporal windows, each of which is then compacted
into a single VLAD, hyper-pooling or VLAC descriptor (referred to as VLAD-GoF, HP-GoF and VLAC-GoF, respectively). A video segment
can then be matched by finding maximum  inner product between its VLAD-GoF,
HP-GoF, or VLAC-GoF descriptor and the corresponding
descriptor from the a GoF in the video. Evidently, the length
of the GoF controls the accuracy of the detection of video segments
within longer videos. In addition, GoFs can also be  overlapping
to allow for better temporal resolution within the matching process.

\subsection{Proposed Vector of Locally Aggregated Centers}

Instead of clustering the local descriptors found within each GoF,
we propose to  cluster the centers of clusters of local descriptors.
The aim is to produce results that are increasingly robust to distortions
that may be found in a typical large video database. Encoding centers
is expected to be more robust to such visual distortions since, compared to local feature descriptors, the centers
of local feature descriptors will vary less when artifacts from processing
 are incurred on video frames.  

Consider $T$ training GoFs stemming from a set of training videos.
From the frames of each $\tau$th GoF ($\tau\in[1, \dots,T]$),
we extract a set of $K_{\tau}$ dense SIFT feature
vectors $\mathfrak{\mathcal{F}}_{\tau}=\left\{ \mathbf{f}_{\tau,1},\,\dots,\mathbf{f}_{\tau,K_{\tau}}\right\}$, each having $F$ dimensions.
From each  $\mathcal{F}_{\tau}$, we calculate  $N$ local feature centers (LFCs)
 $\mathcal{C}_{\tau}=\left\{ \mathbf{c}_{\tau,1},\,\dots,\mathbf{c}_{\tau,N}\right\}$. By concatenating the LFCs for each $\tau$, we acquire the training set of LFCs  $\mathfrak{\mathcal{C}_{\text{train}}}=\left\{ \mathfrak{\mathcal{C}_{1}},\,\dots,\mathfrak{\mathcal{C}_{\mathrm{\mathcal{\mathit{T}}}}}\right\} $.
We then apply a second stage of clustering on  $\mathcal{C}_{\text{train}}$ to generate a set
of $M$   centers of LFCs (CLFCs)  $\mathcal{C}_{\text{enc}}=\left\{ \mathbf{c}_{\text{enc},1},\,...,\,\mathbf{c}_{\text{enc},M}\right\}$,  where each CLFC has $F$ dimensions.

We now consider a test video query $u_{i}$ that contains $G$ GoFs. For
every $g\in[1,\dots, G]$, we extract $K_{g}$ local features
to obtain $\mathfrak{\mathcal{F}}_{g}=\left\{ \mathbf{f}_{g,1},\,...,\,\mathbf{f}_{g,K_{g}}\right\} $.
Then, for every $\mathfrak{\mathcal{F}}_{g}$, we obtain a set of $N$ local feature centers $\mathcal{C}_{g}=\left\{ \mathbf{c}_{g,1},\,...,\,\mathbf{c}_{g,N}\right\}$. Using VLAD we encode each
set of centers $\mathcal{C}_{g}$ with the set of trained centers
$\mathcal{C}_{\text{enc}}$ to generate a vector of locally aggregated
centers (VLAC). Particularly, we first obtain the $F$-dimensional
vector $\mathbf{v}_{g,m}$ for each center $\mathbf{c}_{\text{enc},m}$
in $\mathcal{C}_{\text{enc}}$ by applying
\begin{equation}
\mathbf{v}_{g,m}=\sum_{\forall n:\: Q\left(\mathbf{c}_{g,n}\right)=\mathbf{c}_{\text{enc},m}}\left(\mathbf{c}_{g,n}-\mathbf{c}_{\text{enc},m}\right).\label{eq:VLAD_center_coding-1-1}
\end{equation}
The VLAC for $g$ is then obtained by concatenating $\mathbf{v}_{g,m}$
for all $m\in[1, 2,, \dots,M]$ into a single $1\times MF$-dimensional vector $\mathbf{v}_{g}=\begin{bmatrix}\mathbf{v}_{g,1}, \dots, \mathbf{v}_{g,M}\end{bmatrix}$. We observe that $N$ does not affect the dimension of VLAC, but serves as a control variable for the \emph{coarseness} of the description. After calculating $\mathbf{v}_{g}$ for all $g\in[1,2,,\dots,G]$, we project them on a trained set of $D$ principal eigenvectors to perform  dimensionality reduction. We then concatenate these vectors  to generate
a compact $G\times D$-dimensional vector $\mathbf{v}_{u_{i}}=\begin{bmatrix}\mathbf{v}_{u_{i},1} & \cdots & \mathbf{v}_{u_{i},G}\end{bmatrix}$
for video $u_{i}$. The similarity between two videos $u_1$ and $u_2$  is given
by calculating $s_{u_1,u_2}=\left\langle \mathbf{v}_{u_1},\mathbf{v}_{u_2}\right\rangle $. A threshold is then applied on $s_{u_1,u_2}$ to determine whether
the videos are similar. If two videos contain a different number of
GoFs (e.g.,    $G_{1}$ and $G_2$ GoFs with  $G_{2}>G_{1}$), $s_{u_1,u_2}$ is calculated for all possible alignments $k$
of the vectors $\mathbf{v}_{u_1}$ and $\mathbf{v}_{u_2}$. Finally, the maximum over $k$
is taken to be the similarity score. This can be expressed as
\begin{align}
s_{u_1,u_2} = \max\sum_{g=1}^{G_{1}}&\left\langle \mathbf{v}_{u_1,g},\mathbf{v}_{u_2,g+k}\right\rangle\nonumber\\ &\forall k\,\in\{1,\,2,\,...,(G_2-G_1)\}
\end{align}

Examining the performance of VLAC\ under the experiment of Section \ref{sec:Motivation}, we obtain  $s_{\text{VLAC,clean,noisy}}=0.1131$,  which is more than 13 times higher than   $s_{\text{\textbraceleft VLAD,HP\textbraceright,clean,noisy}}$. We therefore expect the proposed method to be significantly more robust than VLAD and hyper-pooling when assessing video similarity under noisy conditions. However, in order to be suitable for video retrieval, it must also be \textit{discriminative}, i.e., be able to differentiate between \textit{dissimilar} videos that would inherently lead to different features. This is assessed experimentally in the following section.   \

\section{Evaluation of Video Descriptors\label{sec:Evaluation}}
\label{sec:Evaluation}

\subsection{Dataset}

We selected 100 random videos from the Open Video Project (OVP), comprising 1000 minutes of video. Seven types of distortions (Table \ref{tab:Distortions}) were applied
to this footage to examine the performance of VLAD, hyper-pooling (HP)
and VLAC under noise. Training for VLAD, VLAC and HP\ centers was done on different OVP videos from the utilized test material. 

To generate the queries, one-minute video segments were extracted  from each  original videos.
Then, the dataset and query videos were sampled at a  rate of
$\frac{1}{3}$ frames-per-second (fps). The sampling of the query videos, however, is shifted by 0.25 seconds
with respect to the sampling of the videos in the dataset. In this way,  sampling misalignments were also taken into account.  First, we evaluate the similarity detection of the proposed VLAC versus the state-of-the-art VLAD when both are extracted from each sampled frame in the sequence (that is, $\text{GoF}=1$). For VLAD, we set $J=128$, while for VLAC we use $N=128$ and $M=16$. This provides an upper bound on the detection accuracy and assesses the performance of the proposed method versus the standard per-frame VLAD. Next,  the proposed VLAC-GoF is compared against VLAD-GoF\ and HP-GoF, where one descriptor per GoF of 5 frames is derived and the overlap is set to  one frame. Concerning the parameters for each method, we use $J=128$ for VLAD-GoF, $N=256$ and $M=16$ for VLAC-GoF. For HP-GoF,  the number of centers used to encode the first stage VLAD is $\alpha_1 = 128$ and for the second stage $\alpha_2=32$, where we keep $512$ dimensions from the first stage VLAD. 

\begin{table*}[t]
\centering
\begin{tabular}{|c|c|}
\hline 
Distortion & Parameters\tabularnewline
\hline 
\hline 
Scaling & \textbf{FFMPEG:}\texttt{-vf scale=iw/2:-1}\tabularnewline
\hline 
Rotation & \textbf{FFMPEG:}\texttt{-vf rotate} $\frac{\pi}{15}$\tabularnewline
\hline 
Blurring & \textbf{FFMPEG:} \texttt{-vf boxblur 1:2:2}\tabularnewline
\hline 
Compression  & \textbf{FFMPEG:} \texttt{-crf 35}\tabularnewline
\hline 
Gamma Correction & \textbf{FFMPEG:} \texttt{-vf mp 1:1.2:0.5:1.25:1:1:1}\tabularnewline
\hline  
Flicker & \textbf{OpenCV:} Random brightness change (120\%--170\%)\tabularnewline
\hline 
Perspective Change & \textbf{OpenCV} AffineTransform triangle  \texttt{[(0,0),(0.85,0.1),(0,1)]}\tabularnewline
\hline 
\end{tabular}
\caption{Set of  distortions applied to the videos in the database.}
\label{tab:Distortions}
\end{table*}

\begin{figure}
\subfigure[]{
\includegraphics[width=.52\textwidth,trim=.2cm 0.2cm 0.2cm 0.3cm]{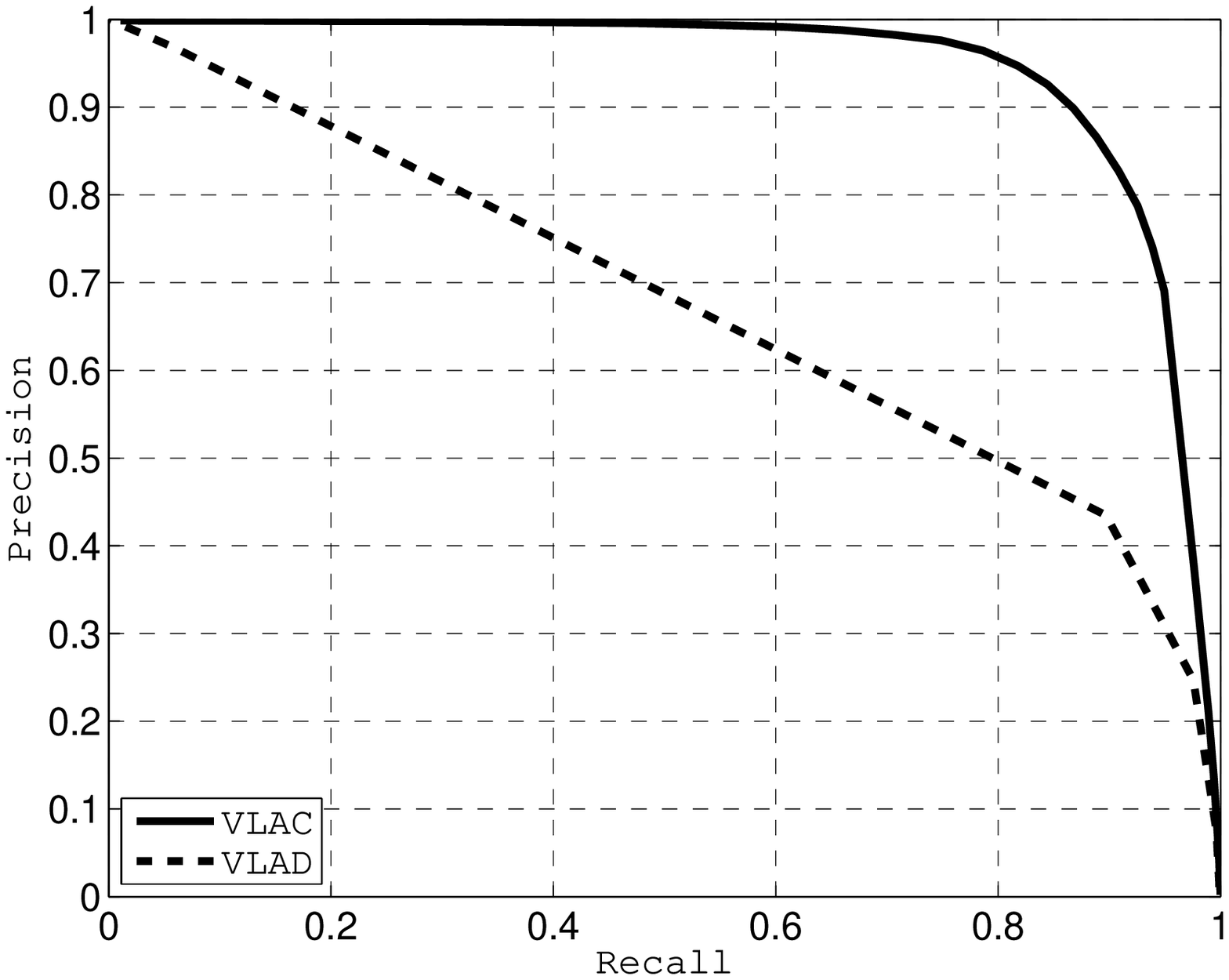}}
\subfigure[]{
\includegraphics[width=.52\textwidth,trim=.2cm 0.2cm 0.2cm 0.3cm]{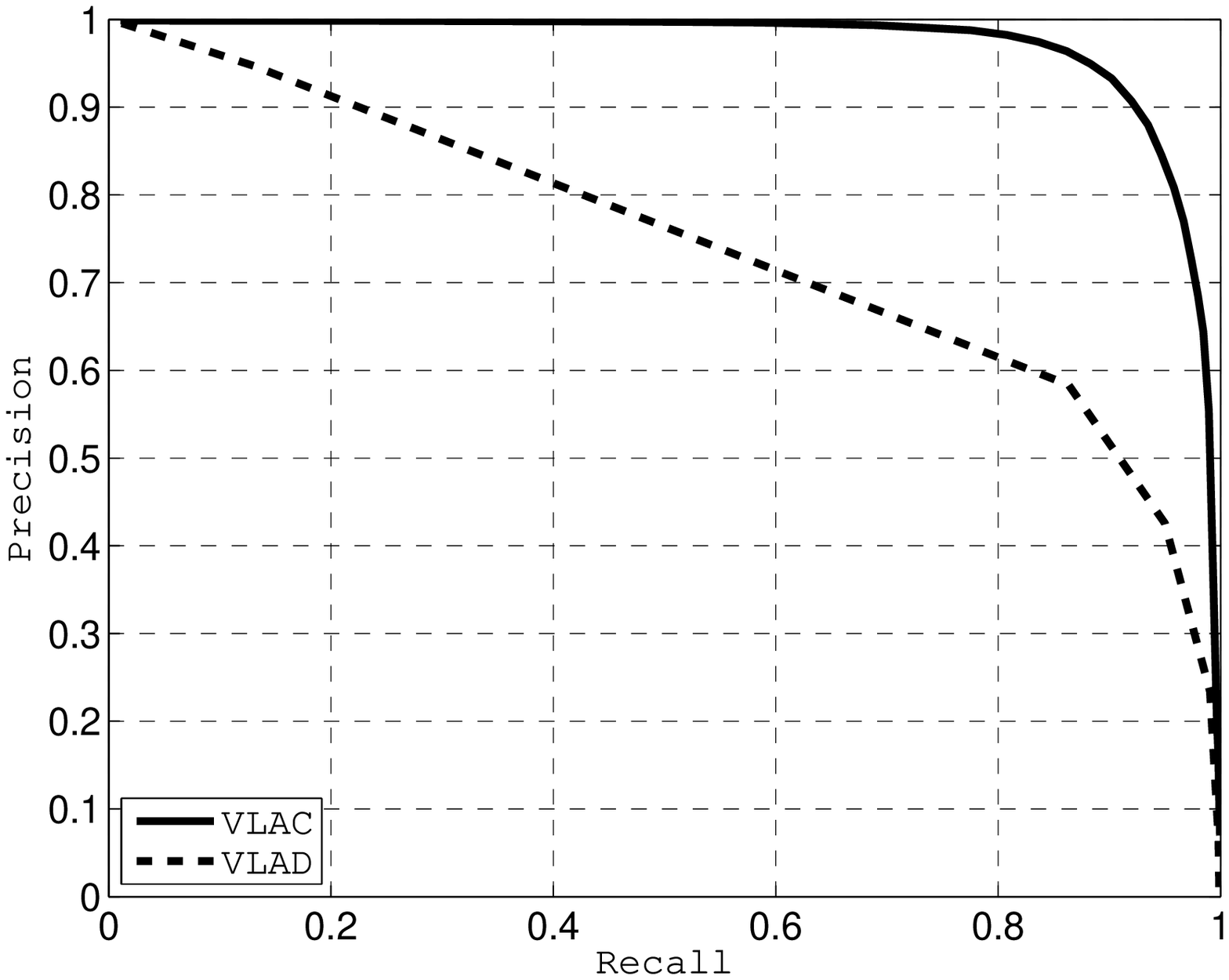}}
\caption{Precision versus recall for VLAD \cite{jegou2012aggregating} and the proposed VLAC,
when extracted per each frame (GoF $=1$); (a) $D=128$ and (b) $D=256$.}
\label{fig:figPerFrame}
\end{figure}

\subsection{Performance and Results}
\begin{figure}
\subfigure[]{
\includegraphics[width=.52\textwidth,trim=.2cm 0.2cm 0.2cm 0.3cm]{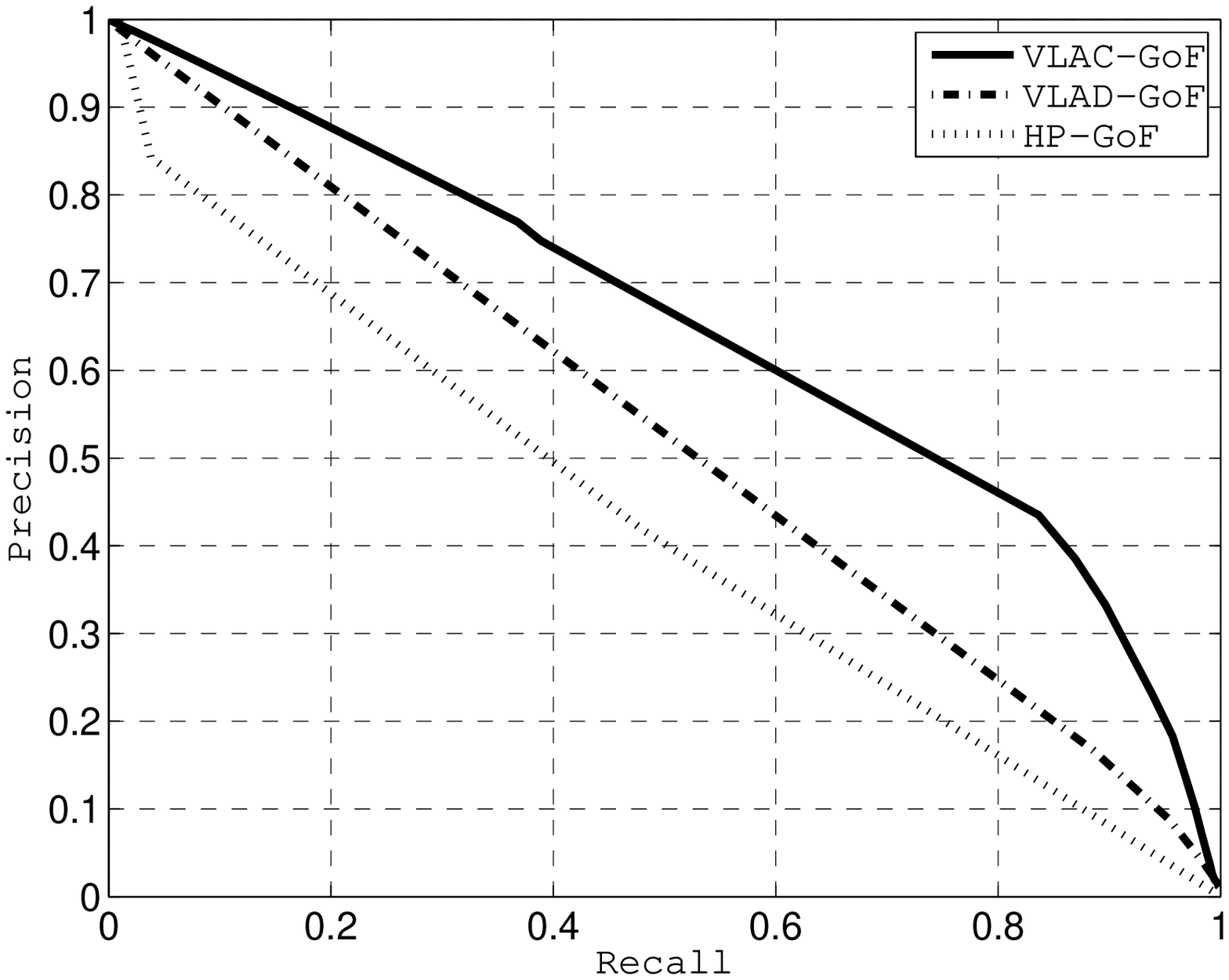}}
\subfigure[]{
\includegraphics[width=.52\textwidth,trim=.2cm 0.2cm 0.2cm 0.3cm]{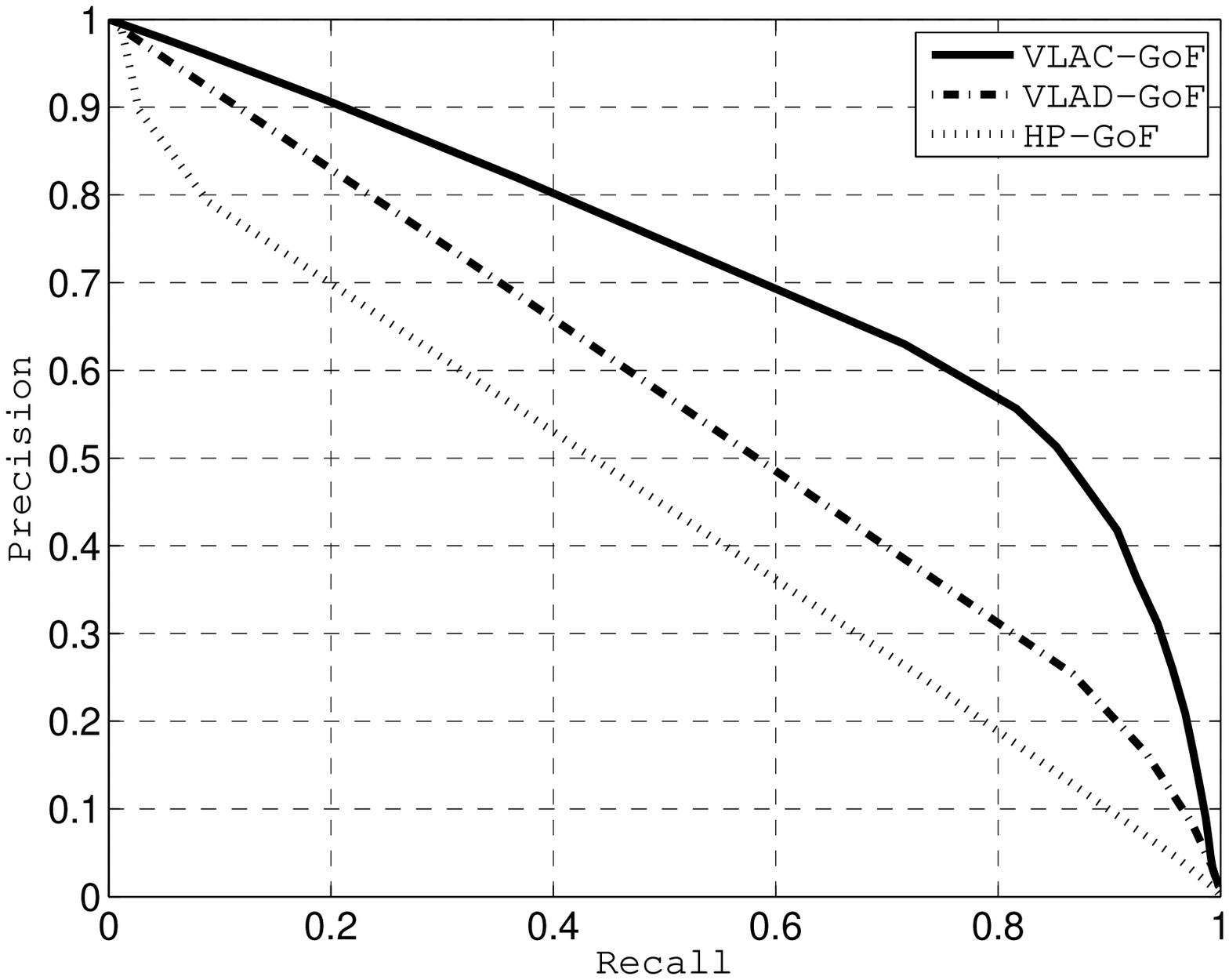}}
\caption{Precision versus recall for VLAD \cite{jegou2012aggregating}, HP \cite{douze2013stable}, and the proposed VLAC under $\text{GoF} = 5$ and the overlap is 1 frame; (a) $D=128$ and (b) $D=256$.}
\label{fig:figGoF}
\end{figure}

Fig. \ref{fig:figPerFrame} depicts the precision versus the recall achieved with the proposed VLAC and the state-of-the-art  VLAD \cite{jegou2012aggregating}, when both descriptors are extracted from each of the frames in the compared video segments.  The results  show that the proposed descriptor offers a substantial detection accuracy improvement compared to VLAD across the entire precision-recall range.  The improved performance of VLAC can be explained by its improved tolerance  to noise, i.e., $s_{\text{VLAC,clean,noisy}}>s_{\text{\textbraceleft VLAD,HP\textbraceright,clean,noisy}}$, which indicates that the principal component projections do not vary substantially after the application of distortions. Therefore, VLAC  retains more information after being projected on its trained principle components. 
Note that the
training videos used to generate the principal
components   did not have any distortions applied on them; this is
to simulate real-life
conditions where we cannot predict the distortions in the dataset.
In addition, all distortions were applied on all videos in the dataset, meaning that higher recall 
reflects higher tolerance to distortions. Same observations can be made from the results in Fig. \ref{fig:figGoF}, where our VLAC is compared against VLAD and hyper-pooling for a GoF of size 5 frames.

Table \ref{tab:mAP}
shows the mean average precision (mAP) for the three compared methods, where $D$ is
the number of dimensions after projection. The results show that, under the same $D$, VLAC improves the mAP by $28.65\%- 38.00\%$ compared to VLAD for frame-by-frame matching and $23.39\%- 26.56\%$ for GoF-based matching. The improvement offered by VLAC-GoF over HP-GoF reaches up to $63.10\%$.

\begin{table}[t]
\centering%
\begin{tabular}{|c|c|c|}
\hline 
 & $D$ & mAP\tabularnewline
\hline
\hline 
VLAD \cite{jegou2012aggregating}  & 256 & 0.7462\tabularnewline
\hline 
 & 128  & 0.6761\tabularnewline
\hline
Proposed VLAC & 256 & 0.9600\tabularnewline
\hline 
 &  128 & 0.9330\tabularnewline
\hline
VLAD-GoF \cite{jegou2012aggregating}  & 256 & 0.5647\tabularnewline
\hline 
 & 128  & 0.5262\tabularnewline
\hline 
Proposed VLAC-GoF & 256 & 0.7147\tabularnewline
\hline 
 &  128 & 0.6493\tabularnewline
\hline 
HP-GoF \cite{douze2013stable} & 256 & 0.4382\tabularnewline
\hline 
 & 128  & 0.4135\tabularnewline
\hline 
\end{tabular}
\caption{Mean Average Precision (mAP) for VLAD \cite{jegou2012aggregating}, HP \cite{douze2013stable} and the proposed
VLAC under: frame-by-frame operation (top two) and GoF-based operation (bottom three).}
\label{tab:mAP}
\end{table}

\section{Conclusion\label{sec:Conclusion}}
\label{sec:Conclusion}
We proposed a novel compact video representation
method based on aggregating local feature centers. Our results show that encoding local feature centers yields significantly better results than simply encoding the features, which are less tolerant to visual distortions commonly found in video databases.  The proposed approach  is therefore suitable for video similarity detection with robustness to visual distortions. The recall-precision results were improved without incurring extra complexity in the signature matching process. Future work will assess the performance of the proposed approach under uncontrolled distortion conditions and even larger datasets.     

\bibliographystyle{IEEEbib}
\nocite{*}

\end{document}